\documentclass[final,5p,times,twocolumn]{elsarticle}

\usepackage{graphicx}
\usepackage{subfig}
\usepackage{amssymb}
\usepackage{amsmath}

\usepackage{lineno}

\usepackage{siunitx}				
\usepackage{caption} 
\usepackage[labelsep=period]{caption}
\usepackage[labelfont=bf]{caption}

\journal{Materials Characterization}

\begin{document}
\begin{frontmatter}

\title{The evolution of precipitate crystal structures in an Al-Mg-Si(-Cu) alloy 
studied by a combined HAADF-STEM and SPED approach}

\author[NTNU]{J.K. Sunde\corref{cor1}}
\ead{jonas.k.sunde@ntnu.no}
\author[SINTEF]{C.D. Marioara}
\author[NTNU]{A.T.J. van~Helvoort}
\author[NTNU]{R. Holmestad}
\address[NTNU]{Department of Physics, Norwegian University of Science and Technology (NTNU), N-7491 Trondheim, Norway}
\address[SINTEF]{SINTEF Industry, N-7465 Trondheim, Norway}
\cortext[cor1]{Corresponding author}
\fntext[fn1]{https://doi.org/10.1016/j.matchar.2018.05.031}
\fntext[fn2]{$\textcopyright$ 2018. This manuscript version is made available under the CC-BY-NC-ND 4.0 license http://creativecommons.org/licenses/by-nc-nd/4.0/}

\begin{abstract}
This work presents a detailed investigation into the effect of a low Cu addition (0.01 at.\%) on precipitation in an Al-0.80Mg-0.85Si alloy during ageing. The precipitate crystal structures were assessed by scanning transmission electron microscopy combined with a novel scanning precession electron diffraction approach, which includes machine learning. The combination of techniques enabled evaluation of the atomic arrangement within individual precipitates, as well as an improved estimate of precipitate phase fractions at each ageing condition, through analysis of a statistically significant number of precipitates. Based on the obtained results, the total amount of solute atoms locked inside precipitates could be approximated. It was shown that even with a Cu content close to impurity levels, the Al-Mg-Si system precipitation was significantly affected with overageing. The principal change was due to a gradually increasing phase fraction of the Cu-containing Q$'$-phase, which eventually was seen to dominate the precipitate structures. The structural overtake could be explained based on a continuous formation of the thermally stable Q$'$-phase, with Cu atomic columns incorporating less Cu than what could potentially be accommodated. 

\end{abstract}


\end{frontmatter}

\section{Introduction}\label{sec:Introduction}
\indent Al-based composites and alloys are prime candidates for \mbox{future} applications due to increasing industrial demands for \mbox{materials} combining properties such as high strength, low weight, good formability and corrosion resistance. They are \mbox{being} used to completely or partially replace, or to be \mbox{combined} with other alloys in a range of applications including \mbox{transport}, construction and packaging \cite{Totten}. The 6xxx series Al(-Mg-Si) \mbox{alloys} form one of the main groups of age-hardening Al \mbox{alloys}. They have demonstrated a combination of \mbox{properties} desirable to the automotive industry \cite{Miller}, and are becoming \mbox{increasingly} utilized in this area \cite{EAA}.  Here, the high strength-to-weight ratio exhibited by these alloys allows for production of lighter \mbox{vehicles} with better fuel efficiency and hence reduced \mbox{emissions}. 

6xxx series Al alloys are characterised by a significant \mbox{increase} in hardness upon short-term thermal ageing. The \mbox{increase} is caused by a large number of nano-sized, semi-coherent and metastable precipitates that form in the Al \mbox{matrix} from the solid solution 
[4,5]. The total alloying addition \mbox{typically} amounts to a few at.\%. Needle-shaped precipitates \mbox{extending} along $\langle$100$\rangle_\text{Al}$ are characteristic of this system, and at peak hardness the microstructure typically comprises a high number density of small $\beta''$-needles \cite{Andersen1}. 
The precipitation \mbox{sequence} in Al-Mg-Si alloys is normally given as 
[7,8]

\begin{align*}
\text{SSSS} &\rightarrow \text{solute clusters} \rightarrow \text{GP-zones }\text{(pre-}\beta'')\\
&\rightarrow \beta''\rightarrow \beta', \text{U1}, \text{U2}, \text{B}' \rightarrow \beta, \text{Si},
\end{align*}

\noindent where SSSS stands for supersaturated solid solution. The \mbox{different} crystal structures can be found elsewhere [9-14].

Many studies have shown that Cu additions in Al-Mg-Si \mbox{alloys} have a positive effect on age-hardening, by providing a higher number density of smaller precipitates [15,16]. 
This has been explained based on density functional theory calculations using two-body interaction energies, where Cu was found to minimize the total energy in the Al-Mg-Si system \cite{Hirosawa}.  
With Cu additions, the 6xxx series precipitation sequence is changed to \cite{Marioara} 

\begin{align*}
\text{SSSS} &\rightarrow \text{solute clusters} \rightarrow \text{GP-zones }\\
&\rightarrow \beta'',\text{L}, \text{S}, \text{C}, 
\beta'\text{-Cu}, \text{Q}' \rightarrow \text{Q}, \text{Si}.
\end{align*}

\noindent See [19-25] 
for the corresponding crystal structures. This implies that the formation of $\beta''$ is suppressed, and other metastable precipitate phases are formed at peak hardness \mbox{conditions}. The different phases can coexist within \mbox{individual} precipitates, forming hybrid precipitate structures. With \mbox{thermal} ageing, this multiphase configuration will undergo \mbox{continuous} changes through diffusive transformations, \mbox{entailing} a complex interplay that occurs on the atomic scale.

The metastable precipitates in the Al-Mg-Si-Cu system are all structurally connected through a mutual network of Si atomic columns. This network exhibits a projected near \mbox{hexagonal} symmetry with $\textit{a} = \textit{b} \approx 4$ \AA, $\textit{c} = n\cdot 4.05$ \AA, $n$ being an integer. Here, the lattice constant \textit{c} is parallel to the needle/rod/lath axis \cite{Marioara}. The precipitates can be described as stacks of atomic columns in $\langle 100 \rangle_{\text{Al}}$ directions adapting to the Al lattice periodicity. They are defined through \mbox{different} arrangements of Al, Mg and Cu atomic columns situated \textit{in-between}, or for certain phases with Cu columns \textit{replacing}, the Si network columns \cite{Saito2}. Mixed element occupancy of atomic columns is also possible. The projected Si network is \mbox{fragmented} in the case of the $\beta''$-phase because of a high \mbox{coherency} with the Al matrix. 

This work focuses on the influence of low amounts of Cu (0.01 at.\%/0.03 wt.\%) because of a reported \mbox{considerable} \mbox{effect} on precipitation when present in higher \mbox{concentrations} \mbox{($\gtrsim$ 0.4 wt.\%)} [23,27-29]. 
Previous studies evaluating the \mbox{effect} of low Cu additions  
looked at lean extruded \mbox{alloys} \mbox{($\lesssim$ 0.1 wt.\%)} [26,30,31]. 
High-resolution transmission \mbox{electron} \mbox{microscopy} (TEM) studies found that $\simeq$ 0.1 wt.\% Cu had \mbox{limited} \mbox{effect} on the Al-Mg-Si system precipitation, merely \mbox{affecting} the \mbox{precipitation} \mbox{kinetics} and number densities. \mbox{However}, the \mbox{precipitates} were \mbox{disordered}, with sub-units of known metastable precipitates in the Al-Mg-Si-Cu system. It was not measured any \mbox{significant} difference in material \mbox{hardness} with $\lesssim$ 0.1 wt.\% Cu. It remains interesting to assess the atomic scale effects of near impurity level Cu additions, and with prolonged ageing times.

\indent Existing methodologies in microstructure quantification of Al alloys by TEM yield information on \mbox{precipitate} \mbox{morphologies} and distributions \cite{Andersen4}, including lengths, \mbox{projected} cross-section areas, number densities and volume fractions \cite{Marioara}. Despite being one of the main obtained \mbox{microstructure} parameters, the total precipitate volume \mbox{fraction} by itself is difficult to clearly relate to any measured \mbox{material} \mbox{property}. What is missing is a statistical assessment of the \mbox{precipitate} phase fractions at each ageing condition, which would enable refinement of the total volume fraction into a \mbox{volume} fraction per phase. Each phase has a different \mbox{chemical} composition, and hence locks different amounts of \mbox{solute} \mbox{elements} inside the structure. The amount and chemical identity of solutes left in solid solution have considerable effect on many \mbox{material} properties, including \mbox{conductivity}, ductility, corrosion and hardness. 

Detailed information regarding coexisting phases and the presence of trace elements inside precipitates can be obtained using e.g. atomically resolved high-angle annular dark-field scanning transmission electron microscopy (HAADF-STEM). Other options include nanobeam diffraction and atom-probe \mbox{tomography} (APT). Despite offering insight to the crystal \mbox{structure} and elemental composition of individual precipitates, the drawback with these approaches is poor statistics, as only a limited number of precipitates can be studied. Especially with larger variations in phases exhibited at each ageing \mbox{condition}, this is a limitation for accurately calculating the precipitate phase fractions. 

The possibility for large variations in Al-Mg-Si-Cu phases existing requires a complementary technique capable of \mbox{probing} a substantial number of precipitates, from which one can infer an overall picture of the frequency of phases. It has been demonstrated that scanning precession electron \mbox{diffraction} (SPED) combined with a machine learning approach can give information on the precipitate phase fractions in the \mbox{microstructure} of Al-Mg-Si-Cu alloys \cite{Sunde}. This \mbox{scanning} technique has the potential of probing a larger number of \mbox{precipitates} than feasible by (S)TEM lattice imaging and APT. The SPED \mbox{technique} was developed further in this work, and combined with HAADF-STEM. This allowed evaluation of both the crystal structure in individual precipitates, as well as a \mbox{quantitative} analysis of \mbox{precipitate} phase fractions using a \mbox{substantial} number of \mbox{precipitates} in each ageing condition. 

The developed technique was employed to achieve the \mbox{following} goals:

\begin{itemize}
\item To obtain an improved understanding of the precipitate crystal structure evolution that occurs during ageing
\item To elaborate the effect of low Cu additions in the Al-Mg-Si system on precipitation in the overageing stage
\end{itemize}

In the first part, the combined HAADF-STEM and SPED methodology used throughout this work is presented along with supporting evidence. In the second part, the obtained results are presented and evaluated.

\section{Experimental procedure}

\subsection{Material and heat treatment}
The material studied was an Al-Mg-Si alloy (6082) with low Cu addition obtained from the materials manufacturer Neuman Aluminium Raufoss (Raufoss Technology). The composition is given in \textbf{Table \ref{tab:composition}}. A cylinder sample (\O\ $\SI{20}{\milli\meter}$, $\SI{10}{\milli\meter}$ height) was cut from an extruded bar and subjected to solution heat treatment at $\SI{540}{\degreeCelsius}$ for $\SI{12}{\minute}$, and then water-quenched to room temperature. $\SI{10}{\minute}$ natural ageing occured before the material was set to artificial ageing at $\SI{180}{\degreeCelsius}$ in an oil bath. The peak hardness condition was obtained after $\SI{3}{\hour}$ ageing, and the additional overaged conditions studied in this work were \mbox{obtained} after $\SI{24}{\hour}$, $1$ week, $2$ weeks and $1$ month total ageing.

\begin{table}[h]
\centering
\caption {Nominal elemental composition of the (6082) Al-Mg-Si(-Cu) alloy studied in this work.}
\label{tab:composition}
\begin{tabular}{l|ccccccc}
\hline
Element & Al & Si & Mg & Cu & Fe & Mn & Cr \\ 
\hline
at.\% & bal. & 0.85 & 0.80 & 0.01 & 0.12 &  0.25 & 0.08 \\
wt.\% & bal. & 0.88 & 0.72 & 0.03 & 0.24 & 0.51 & 0.16 \\
\hline
\end{tabular}
\end{table}

\subsection{TEM specimen preparation}
\indent TEM samples were made by standard electro-polishing. Al discs with $\SI{3}{\milli\meter}$ diameter and thicknesses of approximately $\SI{100}{\micro\meter}$ were electro-polished using a Struers Tenupol-5. The applied electrolyte comprised 1/3 nitric acid (HNO$_{3}$) and 2/3 methanol (CH$_{3}$OH), and was kept at $-\SI{30}{\degreeCelsius}$ to $-\SI{25}{\degreeCelsius}$. The \mbox{applied} \mbox{voltage} was set to $\SI{20}{\volt}$ ($\SI{0.2}{\ampere}$). Prior to SPED and HAADF-STEM experiments, the specimens were cleaned \mbox{using} a \mbox{Fischione} 1020 Plasma Cleaner to reduce the risk of carbon \mbox{contamination} build-up during data acquisition.

\subsection{Electron microscopy}
A JEOL 2100F microscope operated at 200 kV was used for conducting the SPED experiments. SPED was performed \mbox{using} a NanoMEGAS DigiSTAR scan generator retrofitted to the \mbox{instrument}. This system enables the simultaneous scan and \mbox{acquisition} of electron diffraction patterns (DPs) via \mbox{imaging} the phosphor viewing screen of the microscope using an \mbox{externally} mounted 
camera \cite{Moeck}. 

The microscope was operated in nanobeam diffraction mode when performing SPED, using an unprecessed probe diameter of $0.5$, $0.7$ or $\SI{1.0}{\nano\meter}$. The probe semi-convergence angle was $\SI{1.0}{\milli\radian}$. The precession angle employed was $0.5$, $0.7$ or $\SI{1.0}{\degree}$, and the precession frequency was set to 100 Hz. The scan step size was selected as $0.76$, $1.52$ or $\SI{2.28}{\nano\meter}$ depending on the material condition studied. Exposure times used were $20$ or $\SI{40}{\milli\second}$ per pixel. The double-rocking probe required for PED was aligned following the method set out by Barnard \textit{et al.} \cite{Barnard}. A typical dataset comprised $400 \times 400$ PED patterns ($\approx$ 3.2 GB) and required on average $\SI{1}{\hour}$ and $\SI{30}{\min}$ to acquire.

HAADF-STEM images were recorded on a double-corrected JEOL ARM200F microscope operated at $\SI{200}{\kilo\volt}$ using a \mbox{detector} collection angle of \SIrange{42}{178}{\milli\radian}. Conventional \mbox{precipitate} statistics were obtained using a JEOL 2100 \mbox{microscope} ($\SI{200}{\kilo\volt}$). The specimen thickness was \mbox{measured} by \mbox{electron} energy loss spectroscopy (EELS). All TEM \mbox{experiments} were conducted in the $\langle$001$\rangle_\text{Al}$ zone axis of the Al matrix.

\subsection{SPED data analysis}\label{sec:Dataanal}
\begin{figure}[h!]
  \centering
  \includegraphics[width=\columnwidth]{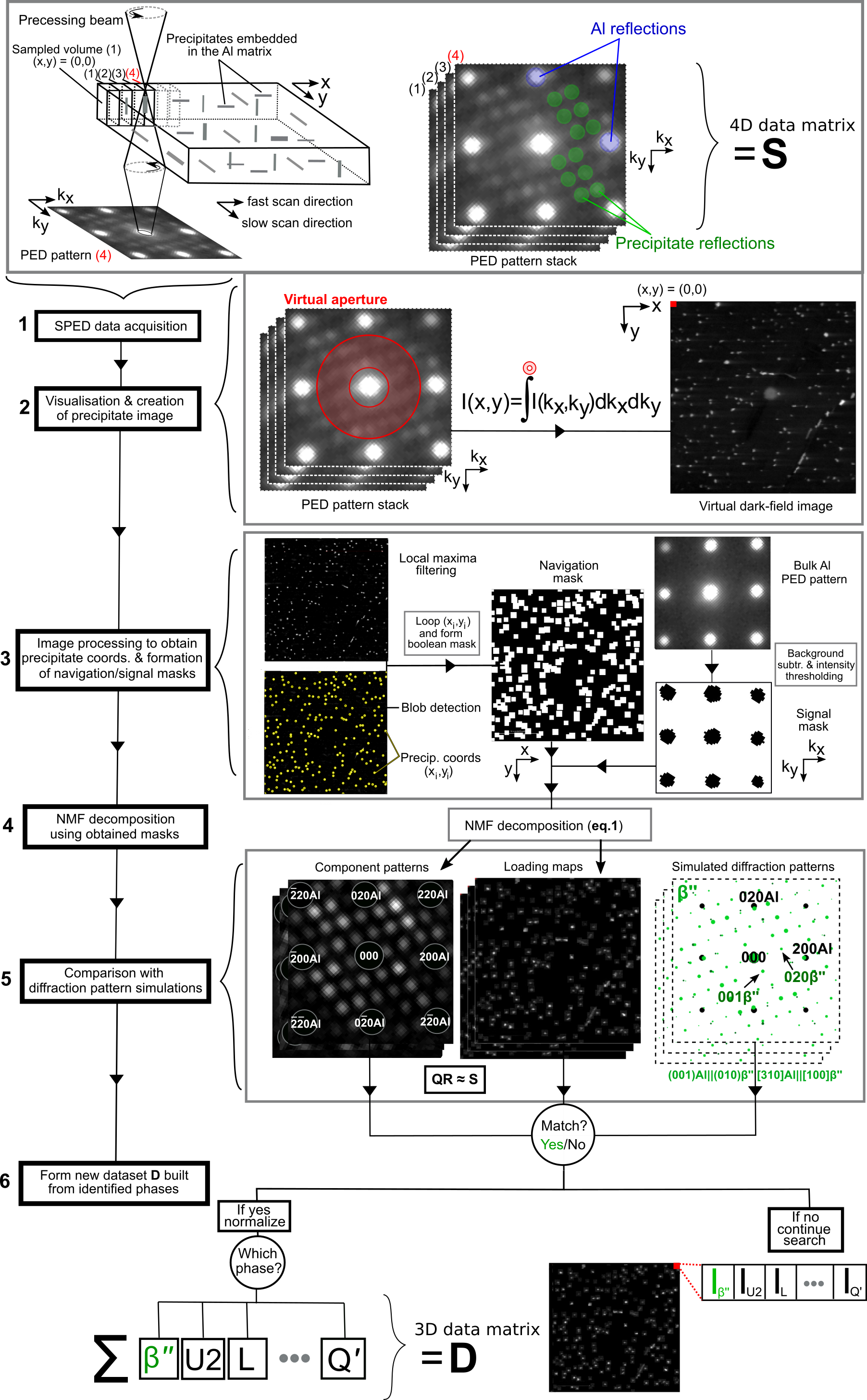}
  \caption{Flow diagram showing the SPED experiment and the data processing applied to the acquired scan data. Details are explained in the associated text.}
\label{fig:DataAnalysis}
\end{figure}

The obtained 4D SPED datasets, comprising a 2D PED \mbox{pattern} at each position of a 2D area scan, were analysed \mbox{using} the HyperSpy \cite{Pena} Python library. The different processing steps are illustrated in \textbf{Fig. \ref{fig:DataAnalysis}}. Initial visualisation of the \mbox{precipitates} present in the scan areas was obtained using \mbox{virtual} dark-field (VDF) images. These were formed by plotting the \mbox{intensity} within a selected sub-set of diffracting pixels, a \textit{virtual aperture}, as a function of probe position. An annular aperture was selected sufficiently large and positioned so as to include diffraction spots from all precipitates lying parallel to the beam direction, verified by TEM images from the same region.

The positions ($x_{i},y_{i}$) of the precipitates were extracted from the VDF images using a \textit{local maxima filtering} and a \textit{blob \mbox{detection}} algorithm based on the Laplacian of Gaussian, both building on Scikit-image project implementations \cite{Scikit}. \mbox{Using} the obtained precipitate coordinates, it was constructed a boolean mask covering the immediate neighbourhood of the precipitate centres ($x_{i},y_{i}$), referred to as a \textit{navigation mask}. In addition, a boolean mask covering the PED pattern Al \mbox{reflections}, a \textit{signal mask}, was formed. The latter was obtained by applying background subtraction to a bulk Al PED pattern followed by intensity thresholding. 

An unsupervised learning approach based on non-negative matrix factorization (NMF) \cite{Lee} was applied to the SPED data, using only the regions of interest marked by the obtained masks, i.e. the pixels covering each precipitate and the pixels falling in-between the Al reflections. The NMF algorithm approximates the 4D dataset, \textbf{S}, by the product of two matrices, \textbf{Q} and \textbf{R}, by minimizing the error function under a positivity requirement

\begin{equation}
\min\limits_{\textbf{Q},\textbf{R}} ||\textbf{S} - \textbf{Q}\textbf{R}||_{\textbf{F}}, \textbf{Q} \geq 0, \textbf{R} \geq 0.
\label{eq:NMF}
\end{equation}

\noindent The inherent condition of positive matrices $\textbf{Q}$ and $\textbf{R}$ ensures more physically meaningful results, being a rational choice in light of the positive nature of PED pattern intensities recorded. The decomposition returned underlying \textit{component patterns} that represent the data, along with associated \textit{loadings} at each real space pixel \cite{Eggeman}. The \textit{loading maps} indicate regions where the component patterns are significant, and resemble simplified dark-field images, see \textbf{Fig. \ref{fig:DataAnalysis}}.

In the case of mixed PED patterns in individual pixels, i.e. patterns composed of reflections from multiple phases, one obtains loading maps with overlapping regions of \mbox{intensity}. Here, the PED patterns have been separated into \mbox{individual} \mbox{components}. This is the key towards unraveling the \mbox{multiple} phase information that can be contained in single pixel PED patterns. This was also attempted by cross-correlation and peak finding approaches, but for the weak precipitate phase \mbox{reflections}, these approaches were deemed inadequate.

The decomposition takes as input the number of \mbox{components} that will approximate the total dataset. Ideally, in the \mbox{regions} \mbox{selected} by the obtained masks, the number of components should equal the number of unique phases present multiplied by the number of allowed orientations, given by their \mbox{orientation} relationship with the Al matrix. This requires the scan area to remain in zone axis orientation. However, this is \mbox{approximately} achieved as long as the bending over the scanned area is smaller than the applied precession angle. \mbox{Furthermore}, the \mbox{integrating} operation of the \mbox{precession} averages out some of the effects from dynamical scattering, and acts to make the PED \mbox{patterns} more 'kinematic-like' \cite{Vincent}. This reduces the \mbox{number} of \mbox{components} needed considerably, and is crucial for machine learning decomposition routines to work. Realistically, due to non-ideal microscope alignments, imperfect masks, disordered phases, strain, dynamical effects and large area tilts, 4-5 times the ideal number of components needed to be included. Through trial-and-error, it was found that 80-120 components were necessary to represent the features of interest in the SPED data. Using a larger number introduced components associated with noise.

After decomposition, the NMF component patterns were manually compared with simulated (kinematic) DPs from known Al-Mg-Si-Cu precipitate phases. Once identified, the associated loading maps provided direct interpretation of real space positions of each phase. The loadings of the identified NMF components were then normalized and summed

\begin{equation*}
I_{\text{i}}(x,y) = \sum_{j}^{N} I_{\text{j}}(x,y).
\end{equation*}

\noindent Here, $(x,y)$ denotes scan pixel coordinate and $I_{\text{j}}$ denotes the \mbox{normalized} loading map for the $j$th component pattern. The \mbox{index} \textit{j} runs over all $N$ \mbox{component} patterns matched to phase $i$, e.g. as a result of the \mbox{different} orientations allowed by the orientation relationship to the Al matrix. Subsequently, it was \mbox{formed} a 3D data matrix \textbf{D} built from all $M$ phases identified, where elements of \textbf{D}, $D_{\text{x,y,i}}= D_{\text{i}}(x,y) = I_{\text{i}}(x,y)$, i.e. denoting the loading value in pixel $(x,y)$ from phase $i$. The net \mbox{result} is a reconstructed and simplified description of the diffraction data, built from loading maps \mbox{associated} with \mbox{component} patterns matched to precipitate phases from the Al-Mg-Si-Cu \mbox{system}.

The data matrix \textbf{D} formed the basis for calculating an \mbox{average} phase fraction for phase $i$, denoted $f_{\text{i}}$, using

\begin{equation}
f_{\text{i}} \approx P^{-1}\cdot\sum_{k}^{P}\left( D_{\text{i}}(x_\text{k},y_\text{k})/ \sum_{i}^{M} D_{\text{i}} (x_\text{k},y_\text{k})\right).
\label{eq:PhaseFract}
\end{equation}

\noindent Here, the sum runs over all precipitate pixels $(x_\text{k},y_\text{k})$, for a total number of $P$ pixels, determined as those pixels which have a significant loading sum $\sum_i D_{\text{i}}(x,y)$.

In summary, a decomposition algorithm based on NMF was applied to the diffraction data from precipitates lying parallel to the beam direction. The decomposition results were matched with phases existing in the material, and combined to form a simplified representation of the SPED scan data. \mbox{Finally}, \mbox{precipitate} phase fractions were estimated using a pixel-based \mbox{calculation} where each precipitate PED pattern has been \mbox{divided} into a sum of the identified phase patterns.
\section{Results and discussions}

\subsection{Conventional TEM analysis}
\begin{figure}[h!]
  \centering
  \includegraphics[width=\columnwidth]{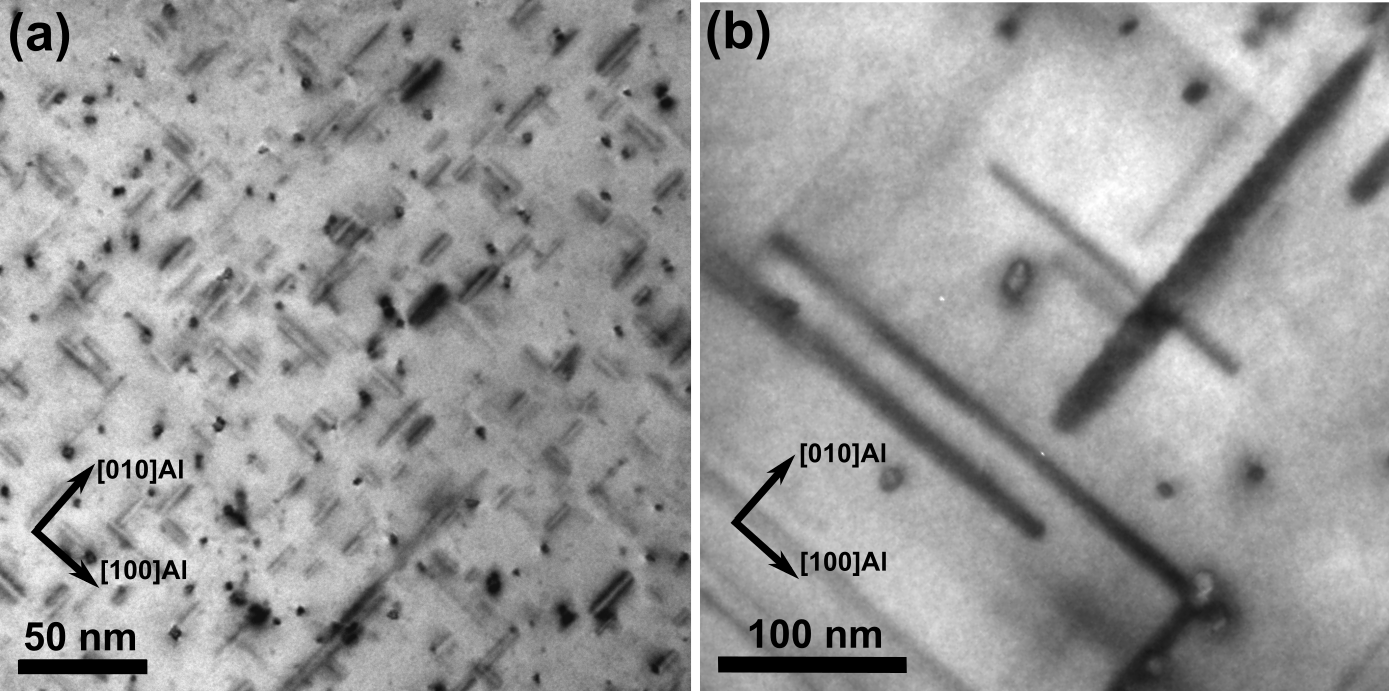}
  \caption{Bright-field image of the (a) $\SI{3}{\hour}$ (peak hardness) and (b) 1 month \mbox{artificial} ageing condition of the Al-Mg-Si(-Cu) alloy studied.}
\label{fig:BFTEM}
\end{figure}

\textbf{Fig. \ref{fig:BFTEM}} shows the large transformation of the Al alloy \mbox{microstructure} when subjected to prolonged ageing at $\SI{180}{\degreeCelsius}$. At $\SI{3}{\hour}$ ageing (peak hardness) the microstructure comprises a high number density of small needles, with an average length of $\SI{15(1)}{\nano\meter}$ and $\SI{13(1)}{\nano\meter\squared}$ average cross-section area. After 1 month total ageing, the precipitates have \mbox{coarsened} \mbox{considerably}, measuring $\SI{352(15)}{\nano\meter}$ in average needle length and $\SI{63(10)}{\nano\meter\squared}$ in average cross-section area. 

\begin{figure*}[h!]
  \centering
  \includegraphics[width=.95\textwidth]{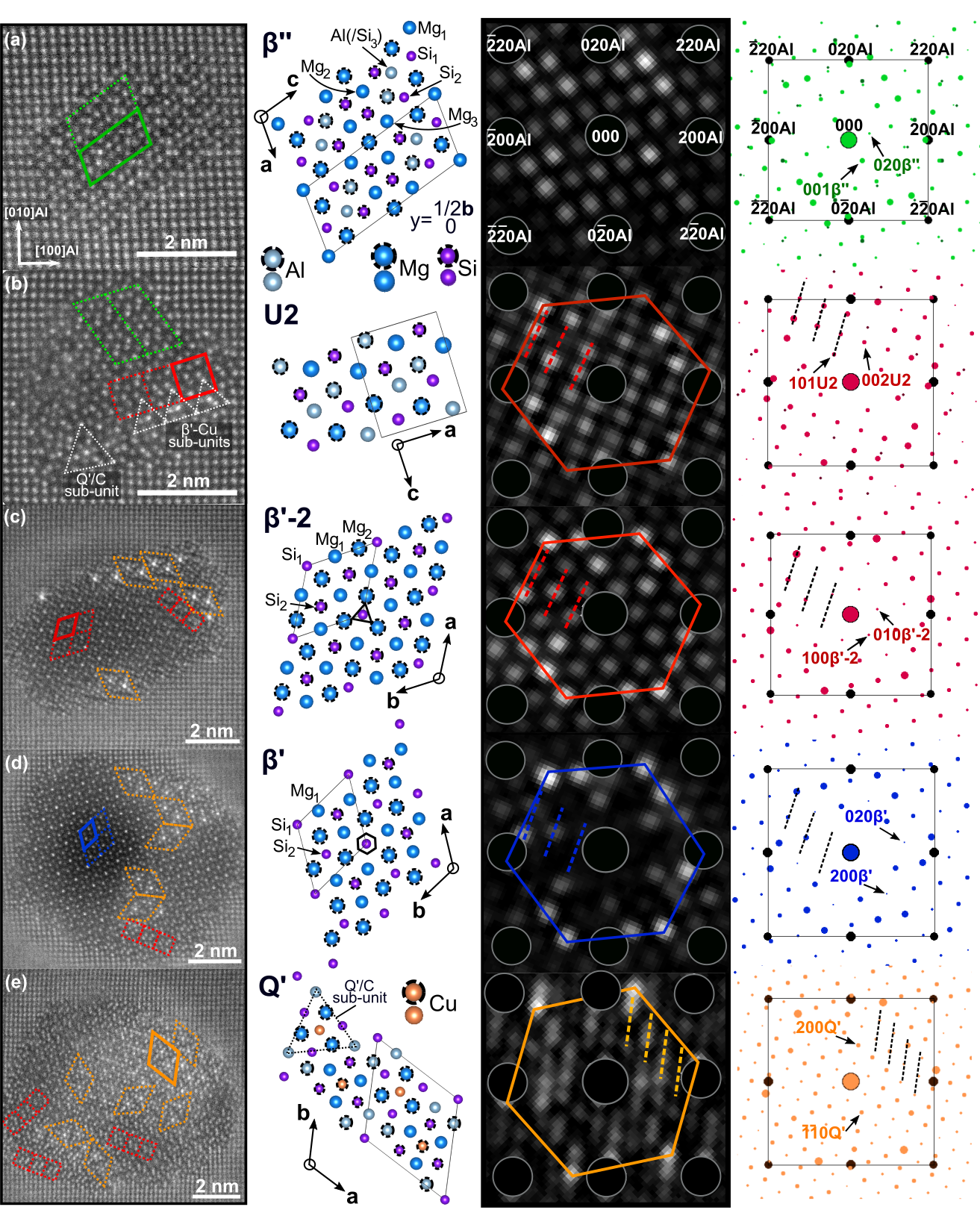}
  \caption{(a-e) HAADF-STEM images of precipitates, precipitate crystal structures based on existing literature, NMF component patterns from SPED data \mbox{decomposition} and matching (kinematic) DP simulations. They are from representative precipitates for each of the ageing conditions studied; $\SI{3}{\hour}$, $\SI{24}{\hour}$, 1 week, \mbox{2 weeks} and 1 month artificial ageing at $\SI{180}{\degreeCelsius}$, respectively. One unit cell of the displayed structures is marked with solid lines, and repeated unit cells are dashed. The hybrid character of the precipitates are indicated, showing several unit cells and sub-units from different phases. Solid and dashed lines in component patterns and matching simulated DPs are guiding aids for the characteristic features of the precipitates.}
\label{fig:HSTEM}
\end{figure*}

HAADF-STEM images of precipitates from the five \mbox{ageing} conditions studied are shown in \textbf{Fig. \ref{fig:HSTEM}}. The \mbox{precipitates} \mbox{presented} were selected from a series of \mbox{images}, and \mbox{evaluated} as representative for each ageing condition. \mbox{Starting} with \mbox{\textbf{Fig. \ref{fig:HSTEM}a}} acquired at $\SI{3}{\hour}$ ageing, the precipitate is seen to be a pure $\beta''$-phase. Some of the atomic columns appear brighter than the other corresponding columns due to partial \mbox{occupancy} of Cu. This is for instance seen at the Al(/Si$_3$) columns in the \mbox{highlighted} unit cell. Cu ($Z_\text{Cu} = 29$) is observed with a higher contrast than the other elements Mg  ($Z_\text{Mg} = 12$), Al ($Z_\text{Al} = 13$), and Si ($Z_\text{Si} = 14$) due to the $Z^{1.7-2.0}$ atom \mbox{column} \mbox{scattering} power at high angles for HAADF-STEM imaging \cite{Nellist}. For some \mbox{precipitates} (not shown), Cu-enriched columns could be seen at the precipitate interface. This has been \mbox{observed} \mbox{previously}, and has been proposed as a \mbox{mechanism} to \mbox{suppress} misfit dislocations \cite{Saito3}.

\textbf{Fig. \ref{fig:HSTEM}b} shows a precipitate at $\SI{24}{\hour}$ ageing. In this condition we see that $\beta''$ is still dominant, but many \mbox{additional} phases are now present, forming a fragmented hybrid structure. Unit cells and sub-units of the U2-phase \mbox{connect} to the unit cells of $\beta''$. At the lower part of the image, sub-units of Cu-containing phases such as Q$'$/C and $\beta'$-Cu are seen, which have Cu columns \mbox{located} \textit{in-between} and \mbox{\textit{replacing}} the Si \mbox{network}, respectively. These phases are conspicuous due to their high Cu content. Several precipitates at this \mbox{condition} showed the same tendency, with a predominantly Al-Mg-Si containing $\beta''$/U2 \mbox{interior}, and Cu-enrichment, as well as \mbox{randomly} positioned sub-units of Cu-containing phases at the precipitate interface. This is similar to earlier results [26,43], 
\mbox{showing} $\beta''$/disordered precipitates at under-aged and peak hardness conditions for alloys with higher Cu content. The positions of Cu enriched columns in the $\beta''$-phase for the $\SI{3}{\hour}$ and $\SI{24}{\hour}$ \mbox{conditions} seem to agree with the findings of Li \textit{et al.} \cite{Li}, which found that Cu \mbox{incompletely} substitutes for Mg$_1$ and Si$_3$ (here, Al) columns. In rare cases, Si$_2$ columns also showed \mbox{partial} Cu occupancy. 

After 1 week ageing (\textbf{Fig. \ref{fig:HSTEM}c}) it is seen that a large \mbox{transformation} of the precipitate structure has occurred. The precipitates have coarsened considerably. The largest phase fraction, often dominating the precipitate interior, is now that of the $\beta'$-phase. Detailed image analysis revealed that there \mbox{exists} a pattern of varying intensity at the Si column \mbox{positions}. Many regions showed a significantly higher intensity at Si$_1$ columns as compared to the Si$_2$ columns, which were more uniform in intensity. These regions are likely associated with a $\beta'$ \mbox{structure} as reported by Vissers \textit{et al.} \cite{Vissers}. There \mbox{evidence} was given for a structure with an \mbox{addition} of 1 extra Si atom per \mbox{3$\times$4.05 \AA} in the \textit{c}-direction at the Si$_1$ columns. This leads to a higher \mbox{column} occupancy, and hence a higher \mbox{intensity} at the Si$_1$ site. However, in other regions the \mbox{difference} \mbox{between} Si$_1$ and Si$_2$ column \mbox{intensities} was less \mbox{distinct}, \mbox{meaning} that the atomic modulation of Si$_1$ columns has not developed. The border between the \mbox{regions} is not very clear, and there exists a tiling of the two. 
 
\begin{figure}[h!]
  \centering
  \includegraphics[width=\columnwidth]{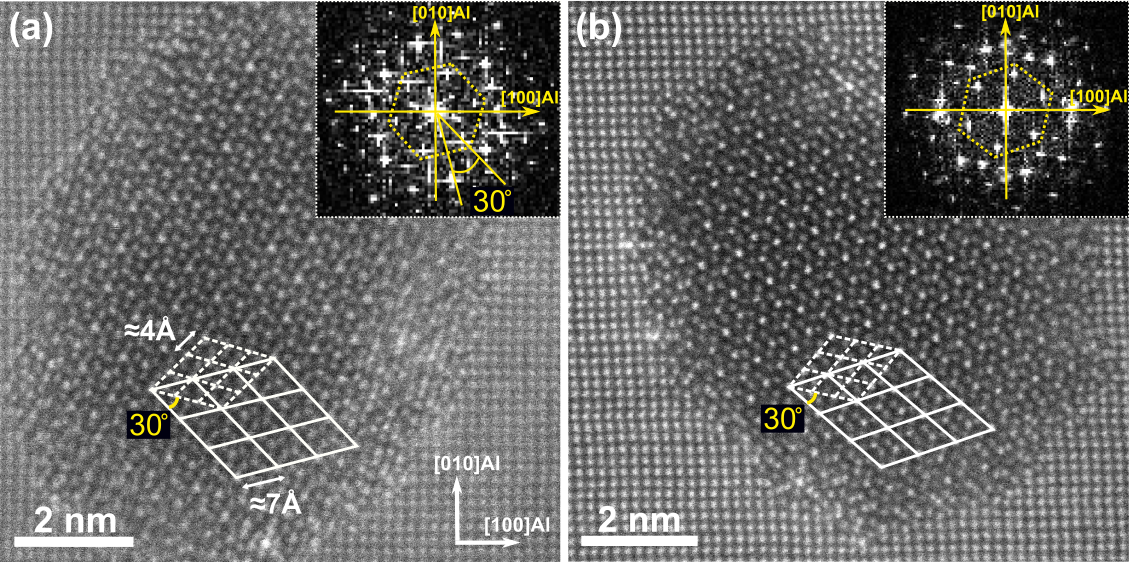}
  \caption{HAADF-STEM images showing (a) $\beta'$-2 and (b) $\beta'$. The insets show fast fourier trasforms (FFTs) of the lattice images. There is a slight, but \mbox{noticeable} difference between the FFTs in (a) and (b) indicating a small \mbox{symmetry} difference between the two $\beta'$ types.}
\label{fig:Betap_diff}
\end{figure}

The difference is more evident in diffraction data, where the presence of added reflections (\textbf{Fig. \ref{fig:HSTEM}c-d}) indicates a structural disparity. Image analysis and DP simulations indicate that the regions without clear modulation of Si$_1$ columns instead have a space group symmetry similar to $\beta'$-Cu(/Ag) \mbox{(P-62m)} \cite{Marioara2}.The difference in space group symmetry gives a \mbox{trigonal}, as opposed to hexagonal symmetry surrounding the corner Si columns in the proper $\beta'$ unit cell (P6$_3$/m), see \textbf{Fig. \ref{fig:HSTEM}c-d}. The change of symmetry relative to proper $\beta'$ seems to be \mbox{related} to the different Si$_1$ column occupancies, which necessarily \mbox{affect} the bonding to the neighbouring atomic columns. The difference is difficult to assess in HAADF-STEM images, as the deviating structure can lie somewhere in-between the two symmetries. However, with the diffraction approach employed throughout this work, the two variants can clearly be separated, and in \mbox{order} to distinguish them, the notation $\beta'$-2 will be used to denote the structure that deviates from proper $\beta'$, see \textbf{Fig. \ref{fig:Betap_diff}}.

Analysis of atomic column Z-contrast indicates that $\beta'$-2 is isostructural with $\beta'$-Cu(/Ag) \cite{Marioara2}, but having Cu replaced by Si, and Al by Mg. This is probably due to the smaller size of the Si atoms as compared to Cu, which allows for more space to be occupied by the larger atom size of Mg. $\beta'$/$\beta'$-2 are the main precipitate constituents in the \mbox{1 week} condition, often linked through a narrow strip of U2 unit cells/sub-units to the Al matrix or phases at the interface. The interface comprises unit cells and sub-units of Al-Mg-Si-Cu phases, most notably Q$'$/C. Compared with the $\SI{24}{\hour}$ condition, the region of Cu-containing phases at the interface has grown in extent. $\beta''$ was not frequently observed in the 1 week condition.

After 2 weeks ageing (\textbf{Fig. \ref{fig:HSTEM}d}) many precipitates show a modest coarsening relative to the 1 week condition. The phases coexisting are similar to the 1 week case. From a \mbox{series} of \mbox{images}, the main observed development of the precipitate \mbox{structure} seems to be the \mbox{progression} of the Q$'$-phase towards the precipitate interior. Several grouped unit cells of Q$'$ can now be seen.

\begin{figure}[h!]
  \centering
  \includegraphics[width=1.\columnwidth]{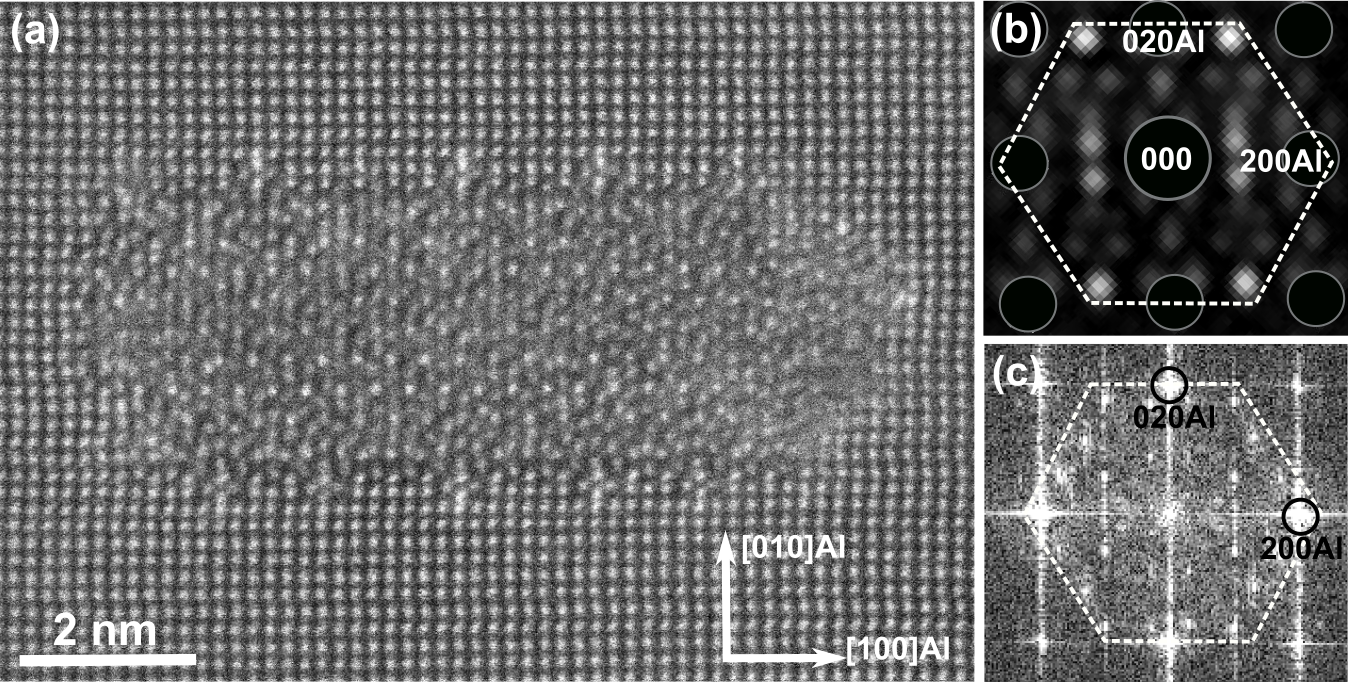}
  \caption{(a) HAADF-STEM image of a L-phase formed on a dislocation line in the $\SI{24}{\hour}$ ageing condition. The associated component pattern (b) shows \mbox{agreement} with the FFT of the lattice image (c). The projected \mbox{hexagonal} Si network is indicated, which is aligned along $\langle$100$\rangle_\text{Al}$ \mbox{directions} for the L-phase.}
\label{fig:Lphase}
\end{figure}

\begin{figure*}[h!]
  \centering
  \includegraphics[width=0.9\textwidth]{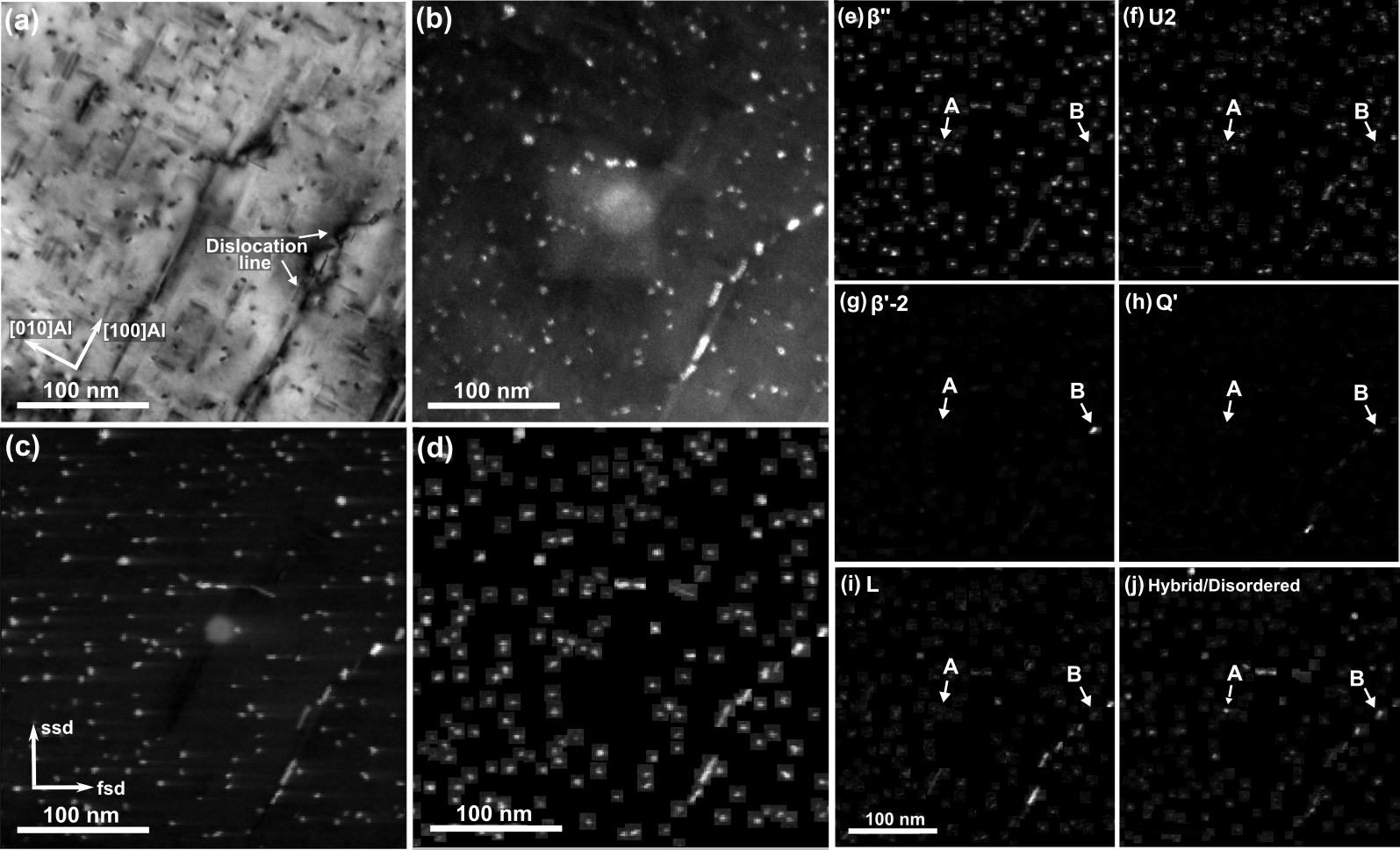}
  \caption{(a) Bright-field and (b) dark-field image acquired in a SPED scan area near the [001]$_\text{Al}$ zone axis. (c) VDF image constructed from precipitate reflections.\\ (d) Reconstruction of the SPED data from NMF component patterns matched with precipitate phases existing. (e-j) Loading maps from each of the identified phases in the SPED scan area. The two precipitates \textbf{A} and \textbf{B} are imaged by HAADF-STEM in  \textbf{Fig. \ref{fig:HybridPrecip}}. \textit{ssd} and \textit{fsd} denotes slow scan- and fast scan direction, respectively.}
\label{fig:ScanArea}
\end{figure*}

At 1 month total ageing (\textbf{Fig. \ref{fig:HSTEM}e}) there is limited further coarsening of precipitates. Similar to the 1 week--2 weeks transition, there seems to be a further development of the Q$'$ takeover of the precipitate structures. Qualitatively, for some precipitates there seems to be a near 50/50 \mbox{division} of $\beta'$/$\beta'$-2 vs. Q$'$. Apart from these phases, there are still lines of U2 unit cells or sub-units linking or 'gluing' the other \mbox{coexisting} phases together. This was the observed behaviour of the U2-phase in all conditions with disordered precipitate \mbox{structures}. Such U2 characteristics have also been observed in the $\beta'$-Cu(/Ag) phase \cite{Marioara2}.

In addition to the phases previously discussed, the L-phase was also observed in all alloy conditions (\textbf{Fig. \ref{fig:Lphase}}). This was however first revealed when the areas were revisited after SPED analysis. Its presence remained low in numbers and stable with ageing, often linked with microstructure defects such as \mbox{dislocation} lines. This is in agreement with previous studies of this phase \cite{Marioara4}. 

In summary, the main observed precipitate phase \mbox{evolution} begins with a $\beta''$ to $\beta'$/$\beta'$-2 transition. Looking at the \mbox{role} of Cu in the initial stages, the atoms likely start out at the $\beta''$/Al interface, where they act to suppress the misfit strain. \mbox{Subsequently}, and in agreement with previous studies, Cu atoms enter the $\beta''$ unit cell preferentially at the Si$_3$ (Al) and Mg$_1$ sites. In addition, sub-units of Cu-containing phases \mbox{(\textbf{Fig. \ref{fig:HSTEM}b})} are formed at the precipitate interface. In the next evolution stage, there occurs a slow movement inwards by the Q$'$-phase, gradually taking over the precipitate interior. Cu has a low diffusion coefficient in the Al matrix, which supports the slow kinetics of Cu-containing phases observed \cite{Du}.  The observed evolution is likely to terminate when there is too few additional Cu atoms left in solid solution to drive the diffusion, and can be followed qualitatively by HAADF-STEM imaging.

\subsection{Assessment of the information content in SPED data}

\begin{figure*}[h!]
  \centering
  \includegraphics[width=.8\textwidth]{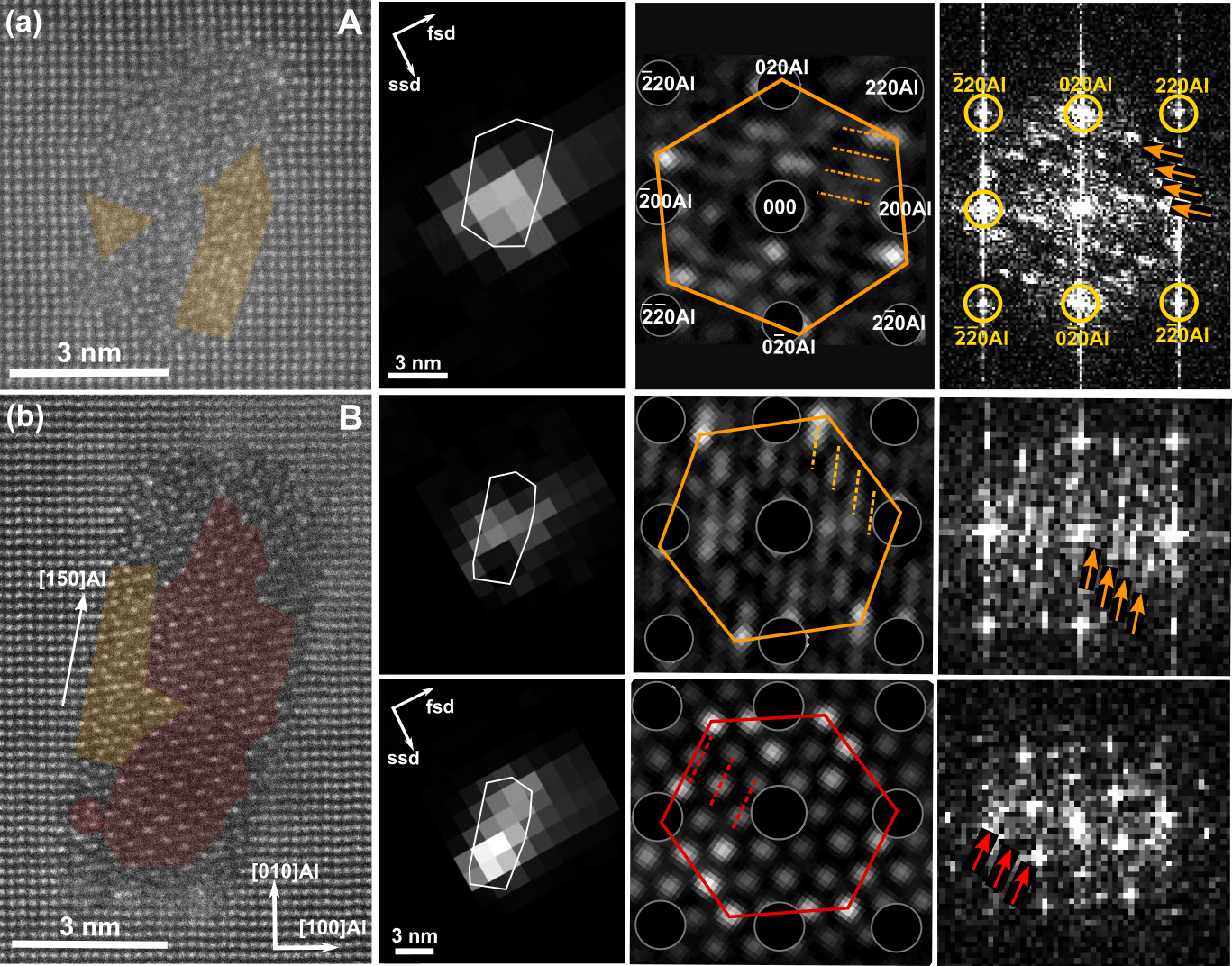}
  \caption{(a) and (b) show HAADF-STEM lattice images and SPED decomposition results for precipitates labeled \textbf{A} and \textbf{B} in \textbf{Fig. \ref{fig:ScanArea}}. The first column shows precipitate lattice images. The second and third columns show the main loading maps and associated component patterns from NMF decomposition, respectively. The fourth column shows FFTs of the lattice image areas highlighted. The red and orange highlighted regions show unit cells/sub-units of $\beta'$-2 and Q$'$, respectively. The outlines of the precipitate cross-sections are overlaid on the loading maps. Characteristic precipitate features in component patterns and image FFTs are indicated.} 
\label{fig:HybridPrecip}
\end{figure*}

In order to obtain statistical support for the observed \mbox{evolution} indicated by atomically resolved imaging, SPED combined with a machine learning approach was pursued. This scanning technique can cover areas over $\SI{1}{\micro\meter\squared}$ in nm-sized steps, which could include 10s to 1000s of precipitate cross-sections depending on the thermomechanical pre-processing applied to the material. This increased sampling forms the \mbox{basis} for improved \mbox{statistics} complementary to conventional high-resolution TEM analysis. For the alloy conditions studied in this work, it was necessary to adapt the step size and hence the scanned area to include a significant number of precipitates, \mbox{especially} for the longest ageing times, as seen from \textbf{Fig. \ref{fig:BFTEM}b}.

The VDF image presented in \textbf{Fig. \ref{fig:ScanArea}c} shows a selection of 188 precipitates, acquired for the $\SI{24}{\hour}$ alloy \mbox{condition}. \mbox{Comparison} with bright-field and dark-field TEM images (\textbf{Fig. \ref{fig:ScanArea}a-b}) \mbox{indicates} that all precipitates (parallel to the beam \mbox{direction}) are detected. The principal loading maps obtained by NMF \mbox{decomposition} and matched with precipitate phases \mbox{existing} are shown in \textbf{Fig. \ref{fig:ScanArea}e-j}. This shows how the technique can be used for phase mapping in the alloy \mbox{microstructure}. \textbf{Fig. \ref{fig:ScanArea}d} shows the sum of the loading maps, i.e. the \mbox{reconstructed} dataset based on the six identified precipitate phase classes ($\beta''$, U2, $\beta'$-2, Q$'$, L, \mbox{hybrid/disordered}). The \mbox{hybrid} class denotes \mbox{component} \mbox{patterns} that comprise a mix of \mbox{individually} \mbox{identifiable} patterns from Al-Mg-Si-Cu phases. Hybrid \mbox{components} are likely to arise in cases where either one of the phases in the mix does not occur in any individual PED pattern, probably caused by the spatial confinement of the phase, which does not extend beyond the scan pixel size. The disordered class denotes component patterns which do not clearly resemble any of the DP simulations, but which contain certain features of \mbox{interest}, such as the Si network hexagon. Comparison with the DF/VDF images (\textbf{Fig. \ref{fig:ScanArea}b-c}) indicates that this reconstructed representation covers the precipitates seen in the area. 

Examples of main component patterns obtained by NMF \mbox{decomposition} for the different alloy conditions are shown in \textbf{Fig. \ref{fig:HSTEM}a-e}. These components show good agreement with \mbox{(kinematic)} simulations of DPs for known Al-Mg-Si-Cu \mbox{precipitate} structures.

It is observed from the loading maps that several \mbox{precipitates} have multiple overlapping pixels of intensity coming from \mbox{different} phases. This shows how phases are revealed down to the sub-precipitate level, i.e. multiple phases within a \mbox{single} precipitate are resolved. The reliability of the \mbox{decomposition} was assessed by correlating the NMF \mbox{results} with atomic \mbox{resolution} images acquired from some of the very same \mbox{precipitates}, e.g. the precipitates labeled \textbf{A} and \textbf{B} in \textbf{Fig. \ref{fig:ScanArea}e-j}, which are shown in \textbf{Fig. \ref{fig:HybridPrecip}a-b}, respectively. 

The precipitate imaged in \textbf{Fig. \ref{fig:HybridPrecip}a} shows a particularly \mbox{disordered} crystal structure, with near absence of complete crystal structure unit cells. Sub-units of phases in the Al-Mg-Si-Cu system can however be discerned. The most significant loading map and its associated component pattern is shown. The \mbox{pattern} shows no clear match with any of the simulated DPs of single phases in \textbf{Fig. \ref{fig:HSTEM}}. However, the feature of \mbox{importance} is the presence of a hexagonal network with \mbox{reciprocal} spacing in agreement with the projected Si network spacing of the Al-Mg-Si(-Cu) system. In addition, the periodicity in diffraction spot rows highlighted agrees with that of the Q$'$-phase, which is also seen from the FFT of the lattice image. The HAADF-STEM image agrees with the presence of a projected \mbox{hexagonal} Si \mbox{network}. HAADF-STEM and SPED both \mbox{conclude} that the precipitate is disordered, but an underlying Si network is present. A few unit cells and sub-units of U2, $\beta'$ and Q$'$ (highlighted in \textbf{Fig. \ref{fig:HybridPrecip}a}) can be seen. The loading map is most \mbox{significant} in the lower region of the precipitate, which agrees with the position of the Q$'$ sub-units. Additional loadings were obtained for this precipitate (not shown). The \mbox{additional} loadings \mbox{individually} described less than 20\% of the precipitate pixels, and their \mbox{associated} component patterns all showed the Si \mbox{network} hexagon.

\begin{figure*}[h!]
  \centering
  \includegraphics[width=.9\textwidth]{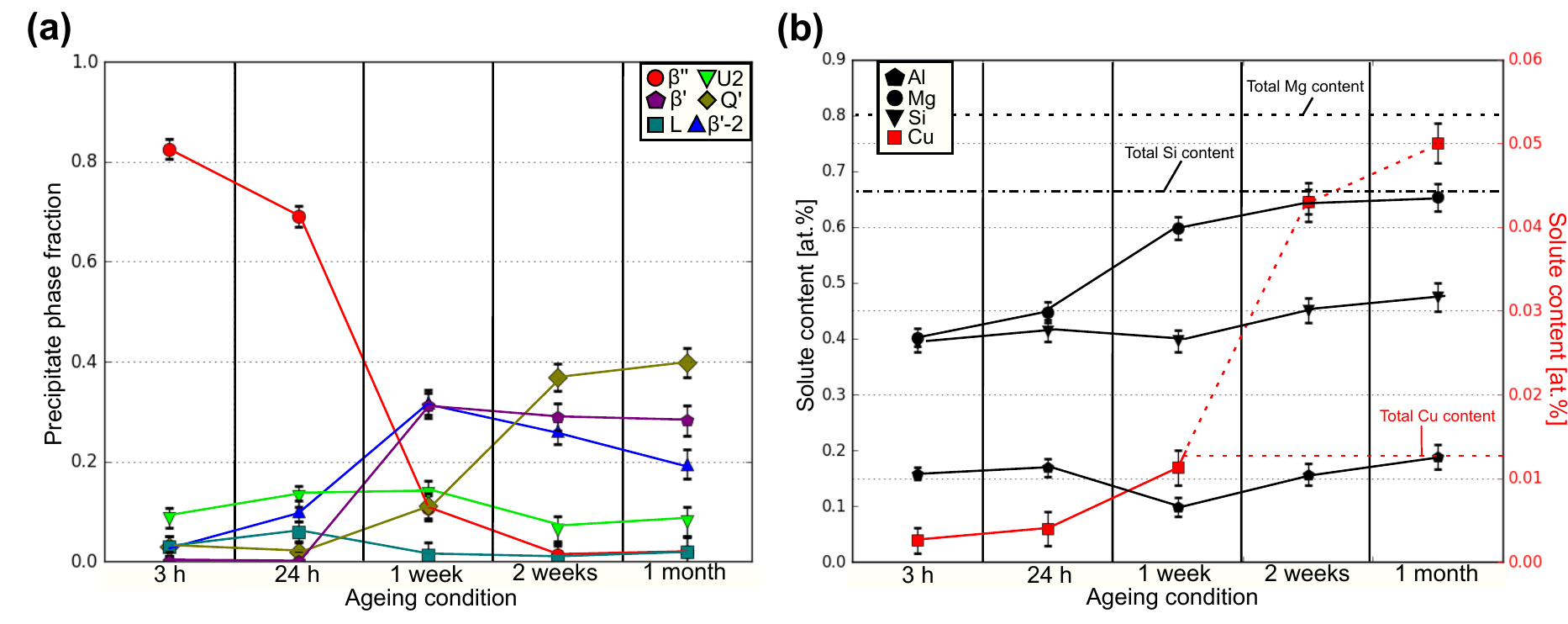}
  \caption{Plots of (a) average precipitate phase fractions (\textbf{eq. \ref{eq:PhaseFract}}) and (b) the total amount of solute atoms locked inside precipitates at each ageing condition studied. The incorporation of Si into dispersoids is taken into account for the marked total Si solute content (dashed-dotted line) using an estimate based on the Alstruc microstructure solidification model \cite{Dons}.} 
\label{fig:PrecipFrac}
\end{figure*}

The second precipitate imaged (\textbf{Fig. \ref{fig:HybridPrecip}b}) shows a \mbox{coarsened} structure located on a dislocation line in the scan area, see \textbf{Fig. \ref{fig:ScanArea}}. Precipitate phases located near dislocation lines were less frequent in this alloy condition (\textbf{Fig. \ref{fig:ScanArea}e-j}), and have likely coarsened as a result of solute segregation or pipe diffusion of elements such as Cu. They hence show a faster ageing \mbox{response} relative to the bulk precipitates. The HAADF-STEM image depicts the presence of a projected hexagonal Si network \mbox{permeating} the structure. The largest phase fraction \mbox{comprising} this precipitate is $\beta'$-2, seen as a bright hexagonal network of Si columns. Additionally, highlighted in the left part of the image is a smaller region of Q$'$ unit cells, whose structure is shown in \textbf{Fig. \ref{fig:HSTEM}e}, having Cu columns situated \textit{in-between} the Si network. Another defining characteristic of this phase is an \mbox{interface} oriented along $\langle 150\rangle$$_\text{Al}$ directions, highlighted in the image. \mbox{Furthermore}, as $\beta'$/$\beta'$-2 does not extend to the Al matrix interface \cite{Teichmann}, there exists a narrow layer of different Al-Mg-Si-Cu phase sub-units. 

The second and third columns of \textbf{Fig. \ref{fig:HybridPrecip}b} show how the SPED analysis provides information about the phases \mbox{coexisting} inside this \mbox{precipitate}. The two main loading maps and associated component \mbox{patterns} obtained are shown. \mbox{Comparison} to DP \mbox{simulations} (\textbf{Fig. \ref{fig:HSTEM}}) identifies the two components as Q$'$ (top) and $\beta'$-2 (bottom), \mbox{respectively}. The \mbox{defining} features are also indicated in the FFTs of the lattice images. The loading maps indicate the phase presence in each pixel. Apart from a visible effect of afterglow in the fast scan \mbox{direction}, the loading maps correspond \mbox{reasonably} well with the HAADF-STEM image.

The main strength of the SPED technique lies in the \mbox{ability} to scan large areas containing a substantial number of \mbox{precipitates}, while still preserving high spatial resolution. This forms the \mbox{basis} for objective evaluation of the precipitate phase \mbox{occurrence} using a representative distribution of precipitates. \mbox{Generalizing} from a series of 10-20 HAADF-STEM images can give a misleading conclusion on precipitate phase \mbox{fractions} in the alloy microstructure, especially when the variations \mbox{between} precipitate phases exhibited are relatively large. There is a risk of selectively probing larger, Cu-containing precipitates due to the strong Z-contrast dependence of HAADF-STEM. Adding to this point, a poor depth of focus leads to further selective imaging of longer, coarsened precipitate needles that reach the specimen surface. 

Besides the risk of selectively probing precipitates, rare cases such as the L-phase (\textbf{Fig. \ref{fig:Lphase}}) or the Q$'$/$\beta'$-2 phase (\textbf{Fig. \ref{fig:HybridPrecip}b}) might be missed in a HAADF-STEM analysis. These \mbox{images} were acquired after analysing the SPED scan data, where the precipitate PED patterns from the dislocation line \mbox{revealed} the presence of rare phases in this alloy condition (\textbf{Fig. \ref{fig:ScanArea}e-j}). It \mbox{becomes} clear that a scanning technique \mbox{capable} of \mbox{covering} large areas will give a more general picture of phase \mbox{distributions}, and hence an improved estimate of precipitate phase fractions in the alloy microstructure. The SPED scan data presented in this work covers precipitate numbers ranging from 520 ($\SI{3}{\hour}$ condition) to 101 (1 month condition). The scan areas and hence the number of precipitates can be further upscaled \cite{Sunde}, and comes at a relatively low cost in increased analysis time. 

The main limiting factors with SPED are currently due to the acquisition system. The diffraction from particularly small \mbox{precipitates} fall close to the noise threshold. In addition, \mbox{adverse} effects of afterglow is observed, see the second \mbox{column} of \mbox{\textbf{Fig. \ref{fig:HybridPrecip}}}. This has an effect on the recorded PED patterns in the fast scan direction, and for large and strongly diffracting \mbox{precipitates} this effect is pronounced. It is however unlikely that the main characteristic of a precipitate is miscalculated due to this effect, as the relative intensities remain similar. There are technical solutions coming to solve these issues.

\subsection{Quantification of precipitate phase fractions}
\textbf{Fig. \ref{fig:PrecipFrac}} summarizes the main findings from the SPED \mbox{analysis} for the five ageing conditions and the total of \mbox{approximately} 1200 precipitates identified. 
\textbf{Fig. \ref{fig:PrecipFrac}a} shows the \mbox{average} \mbox{precipitate} phase fractions estimated from a \mbox{representative} \mbox{distribution} of precipitates in each ageing condition. 
Errors in individual precipitate phase fractions were \mbox{estimated} based on comparison with high-resolution images, which takes into \mbox{account} the adverse effects introduced by the SPED \mbox{acquisition} system. This error was subsequently scaled \mbox{according} to the number of precipitates in the recorded SPED scan, $N$, \mbox{assuming} linearly independent measurements of phase fractions for each precipitate. The plot showcases the main precipitate phase \mbox{evolution} that occurs during ageing.

In line with the HAADF-STEM results, the presence of $\beta''$ is seen to drop quickly with increased ageing time. At the 1 week condition it has dropped to 10\%. Furthermore, it \mbox{confirms} the distinct transformation of precipitate crystal structures from the $\SI{24}{\hour}$ to the 1 week condition. The main phases present changes from $\beta''$ and U2 to $\beta'$/$\beta'$-2, and it is interesting to note that the onset of $\beta'$-2 formation seems to occur before $\beta'$. The 1 week condition also shows an increasing Q$'$ phase fraction ($\approx$11\%). At this condition Q$'$ is predominantly observed inside bulk \mbox{precipitates}, unlike earlier conditions where it was mainly detected near microstructure defects.

\begin{figure}[h!]
  \centering
  \includegraphics[width=1.\columnwidth]{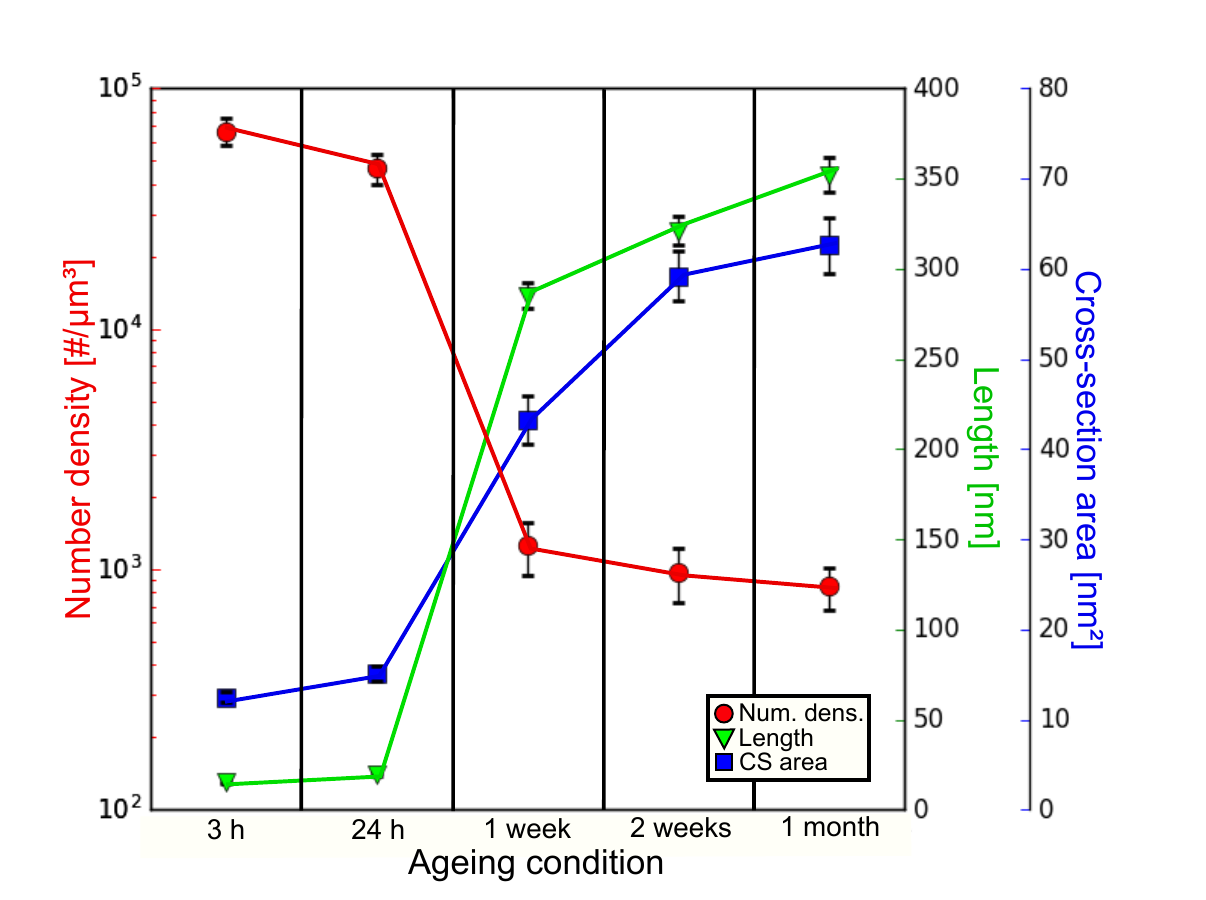}
  \caption{Plot of the precipitate statistics at each ageing condition studied.}
\label{fig:PrecipStat}
\end{figure}

With further ageing there are different developments for the three main phase fractions $\beta'$, $\beta'$-2 and Q$'$. The $\beta'$-2 phase \mbox{fraction} is seen to gradually decrease from 1 week to 1 month ageing, whereas $\beta'$ stays nearly constant at about 30\%. As was observed in HAADF-STEM images, the Q$'$ phase \mbox{fraction} \mbox{increases} steadily as a function of ageing time, and is seen from the plot to reach a phase fraction of $\approx$ 40\% at 1 month \mbox{ageing}. The precipitate cross-section area exhibits a \mbox{modest} growth from 2 weeks to 1 month ageing, seen from \textbf{Fig. \ref{fig:PrecipStat}}. \mbox{Interpretation} of the observed $\beta'$/$\beta'$-2 evolution suggests that in order for the $\beta'$ phase fraction to stay nearly constant, with \mbox{limited} cross-section area increase and with a gradually \mbox{increasing} fraction of Q$'$, both or either one of two effects are occurring:

\begin{itemize}
\item $\beta'$-2 transforms into $\beta'$ with increased ageing time
\item $\beta'$ remains stable, and the formation of Q$'$ occurs \mbox{preferentially} at $\beta'$-2 regions
\end{itemize}

As previously discussed, the observed evolution is likely to end when there are too few Cu atoms left in solid \mbox{solution}. Using the obtained phase fractions together with \mbox{measured} \mbox{precipitate} statistics, one can estimate the amount of solute atoms locked inside precipitates. The different phases have \mbox{different} chemical compositions and unit cell parameters, see \textbf{Table \ref{tab:PrecipStruct}}. The compositions listed are in the case of the $\beta''$-phase based on an average composition using three structures \cite{Hasting}. For the disordered L-phase, the composition was \mbox{estimated} \mbox{using} HAADF-STEM images, assuming a C-phase atom \mbox{column} density \cite{Torsaeter}. $\beta'$-2 was assumed to have the same average \mbox{composition} as $\beta'$.

\begin{table}[h!]
	\centering 
    \caption{\label{tabone}Average composition and crystal lattice parameters of the precipitate phases found in the Al-Mg-Si(-Cu) alloy studied.} 
    \begin{tabular}{c c c c}
    	\hline
    	Precip. & Chemical & Lattice & Ref.\\
    	phase & composition & parameters [\AA] &\\
        \hline
         &  & $a=15.16,$&\\
        $\beta''$ & Al$_5$Mg$_{14}$Si$_{14}$ & $b=4.05,\ \beta = 105.3^{\circ}$,& \cite{Hasting}\\
         &  & $c=6.74$ &\\
        $\beta'$/$\beta'$-2 & Mg$_6$Si$_{3.33}$ & $a=b=7.15,$& \cite{Vissers}\\
        &  & $c=4.05,\ \gamma = 120^{\circ}$&\\
        &  & $a=6.75,$&\\
        U2 & AlMgSi & $b=4.05,$ & \cite{Andersen2}\\
        &  & $c=7.94$&\\
        &  & $a=10.32,$&\\
        L & Al$_{20}$Mg$_{44}$Si$_{33}$Cu$_2$ & $b=4.05,\ \beta = 100.9^{\circ}$, & \cite{Torsaeter}\\
        &  & $c=8.10$&\\
        Q$'$ & Al$_6$Mg$_6$Si$_7$Cu$_2$ & $a=b=10.32,$& \cite{Wenner}\\
        &  & $c=4.05,\ \gamma = 120^{\circ}$&\\
        \hline
    \end{tabular}
\label{tab:PrecipStruct}
\end{table}

The precipitate phase fractions were first used to convert a measured overall volume fraction (VF) into a \mbox{volume} fraction per phase, VF$_\text{i}$. Subsequently, the \mbox{precipitate} phase unit cell \mbox{parameters} and composition were used to \mbox{convert} a volume fraction per phase into a solute fraction per phase (SF$_\text{i}$). This was calculated using a conversion factor, $k$, relating the two \mbox{parameters}

\begin{equation*}
\text{VF}_\text{i} = k\cdot \text{SF}_\text{i}.
\end{equation*}

The conversion factor is defined as 

\begin{equation*}
k = \left(\frac{\text{equiv. }\#\text{ Al atoms in V}_\text{UC,i}}{\#\text{ atoms in UC}_\text{i}}\right) \cdot \left(\frac{\#\text{ atoms in UC}_\text{i}}{\#\text{ non-Al atoms in UC}_\text{i}}\right),
\end{equation*}

where UC$_\text{i}$ denotes the unit cell of phase $i$ and V$_\text{UC,i}$ is its \mbox{corresponding} volume. Finally, the solute fractions were summed up to give a total solute content locked inside \mbox{precipitates}. 

The results are presented in  \textbf{Fig. \ref{fig:PrecipFrac}b}, which shows the \mbox{development} of total solute content of main elements locked in precipitates during ageing. Besides a dip in the Al solute \mbox{fraction}, and a constant (or small decrease) in the solute \mbox{fraction} of Si in the $\SI{24}{\hour}$ to 1 week transition, Al, Mg, Si and Cu \mbox{content} in precipitates gradually increase with ageing. The noted \mbox{exception} for Al and Si solute fractions at $\SI{24}{\hour}$ to 1 week ageing is due to the formation of $\beta'$/$\beta'$-2. Unlike the other phases observed, which have Si:Mg ratios close to one, $\beta'$/$\beta'$-2 show the largest deviation with an average composition of Mg$_6$Si$_{3.33}$. This necessarily leads to a larger incorporation of Mg than Si in the precipitate structure, which is dissolved back into the solid solution. These structures also contain no Al, which explains the decrease of the Al solute fraction. The combined total solute content in precipitates is monotonously increasing with increased ageing time, meaning that the orders of magnitude decline in number density seen from \textbf{Fig. \ref{fig:PrecipStat}} is compensated by the diffusive growth of large precipitates.

The Cu solute content remains low in the $\SI{3}{\hour}$ to $\SI{24}{\hour}$ \mbox{transition}. At this stage, the Cu content in \mbox{precipitates} is mainly attributed to the L and Q$'$-phase having formed near \mbox{microstructure} defects, such as dislocation lines. HAADF-STEM images show that there additionally exists an \mbox{incorporation} of Cu in e.g. the $\beta''$ structure and a Cu-enrichment at the precipitate interface. This contributes \mbox{however} little to the total Cu content inside precipitates.

\begin{figure}[h!]
  \centering
  \includegraphics[width=.75\columnwidth]{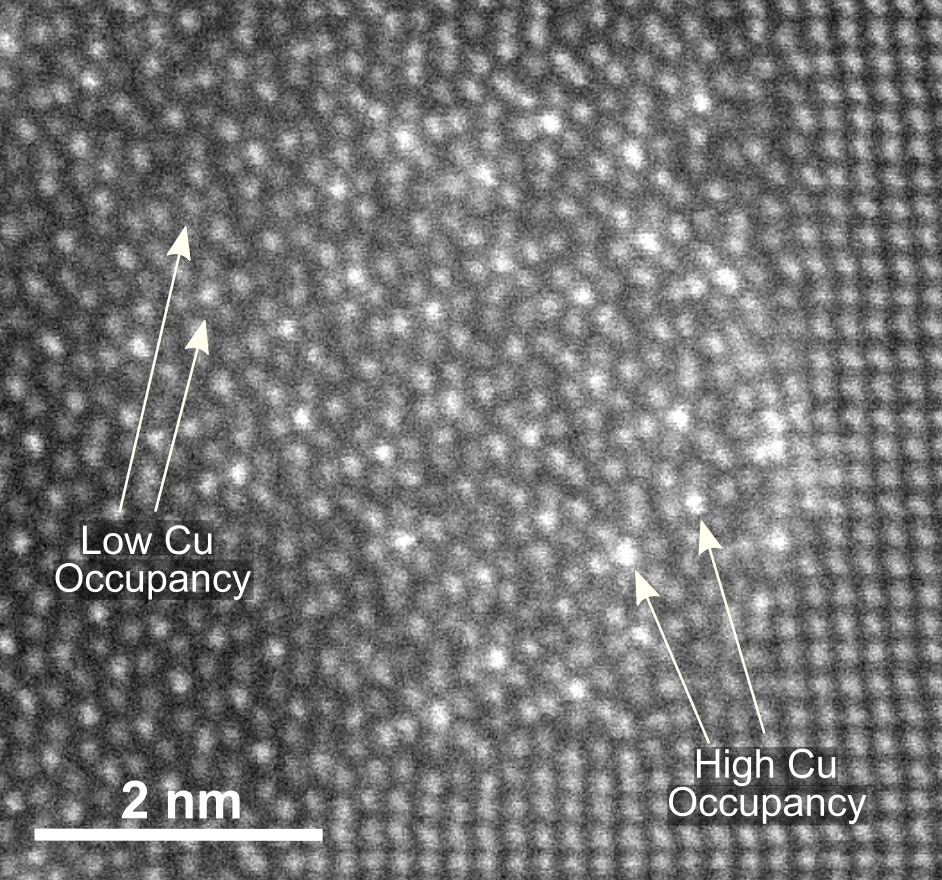}  
  \caption{HAADF-STEM image of a precipitate containing Q$'$. The varying intensity in the indicated atomic columns implies different Cu occupancies.}
\label{fig:CuOcc}
\end{figure}

At the 1 week condition there is a significantly increased phase fraction of Q$'$, which leads to a higher content of Cu in precipitates. This further increases rapidly with ageing, due to a continuously increasing Q$'$ phase fraction. At 1 month ageing it is obtained a total solute fraction of 0.05 at.\% Cu, which far surpasses the total Cu content of 0.013 at.\% in the alloy, marked in \textbf{Fig. \ref{fig:PrecipFrac}b}. The reason is that the calculation assumes a maximum Cu incorporation in the Q$'$ structure of \mbox{$\approx$ 10\%}, which is based on the stated composition Al$_6$Mg$_6$Si$_7$Cu$_2$. \mbox{However}, it is observed here that there exists only a partial occupancy of Cu in the Cu atomic columns of the Q$'$ structure. This is clearly \mbox{seen} in HAADF-STEM images, see \textbf{Fig. \ref{fig:CuOcc}}. The Cu \mbox{content} instead varies between 0-10\%, leading to an \mbox{overestimate} using the maximum value. 

What can be concluded is that the Q$'$ phase fraction near the \mbox{1 week} condition has the \textit{potential} to incorporate nearly all the Cu in the material, being the only phase with a \mbox{significant} Cu content. However, instead the phase continues to form, with less Cu than what can be accommodated in the \mbox{structure}. This explains how even a Cu content almost down to impurity \mbox{levels} can considerably affect the Al-Mg-Si system \mbox{precipitation}. The Q$'$ structure is more thermally stable than $\beta'$/$\beta'$-2, \mbox{evident} by its gradual dominance of the precipitate structure with \mbox{overageing}. Several HAADF-STEM images showed that the Cu occupancy was higher at the precipitate interface, where the Q$'$ unit cells initially are formed. This \mbox{supports} the indication of an \mbox{inwards} movement into the precipitate structures by the Q$'$-phase, which occurs through a slow diffusive process. The slow \mbox{transformation} is due to the low \mbox{diffusivity} of Cu in the Al \mbox{matrix} \cite{Du}.

Different from other phases, the U2 and L phase fractions \mbox{remain} stable ($\lesssim$ 16\%) throughout the total ageing process. This is in agreement with HAADF-STEM lattice images, where the U2-phase is seen to form lines of unit cells in all conditions, forming the 'glue' between the other phases, or the interface to the Al matrix. Among the Al-Mg-Si \mbox{containing} phases, this is the phase that remains most stable in number with ageing. 

The L-phase also remains stable and low in numbers. This is thought to be due to the similarity in the number of \mbox{microstructure} defects in the samples studied, such as grain boundaries or dislocation lines, to which the L-phase is \mbox{associated}. This is clear from the presented $\SI{24}{\hour}$ condition SPED scan (\textbf{Fig. \ref{fig:ScanArea}i}), where a dislocation line contains nearly all the L-phases in this area. The L-phase is known to have high thermal stability, and forms in the bulk at higher alloy Cu contents \cite{Marioara4}.

\section{Conclusions}
This work has demonstrated that combining HAADF-STEM and SPED resulted in a more accurate evaluation and \mbox{quantification} of the precipitate crystal structure evolution that occurs during ageing of a low Cu content Al-Mg-Si alloy. NMF decomposition of SPED data could be used to resolve phases within individual precipitates, and allowed for a \mbox{statistically} sound number of precipitates to be included in the analysis, \mbox{beyond} what is \mbox{achievable} by other \mbox{techniques}. The results \mbox{enabled} an \mbox{estimate} of \mbox{precipitate} phase fractions at each \mbox{ageing} condition, which were used to \mbox{approximate} the total solute \mbox{content} locked inside precipitates throughout the total ageing process.

At peak hardness, the precipitates were \mbox{predominantly} \mbox{observed} as pure $\beta''$. Upon further heat \mbox{treatment}, the \mbox{precipitates} evolved into complex hybrid structures, \mbox{primarily} comprising Al-Mg-Si phases like $\beta''$ and U2 in the precipitate interior. 
Cu-enriched columns and sub-units of Cu-containing phases like Q$'$/C and $\beta'$-Cu existed at the \mbox{precipitate} interface. With increased ageing the precipitates coarsened \mbox{substantially}, and exhibited a projected \mbox{hexagonal} Si \mbox{network} with $\beta'$/$\beta'$-2 in the precipitate interior. Unit cells/sub-units of Cu-containing phases were still \mbox{confined} to the precipitate interface, but had grown larger in \mbox{extent}. \mbox{Beyond} this point it was seen a slow, but gradual \mbox{progression} inwards into the precipitates by the Cu-containing Q$'$-phase, with Cu atomic columns incorporating less Cu atoms than what could potentially be accommodated. Q$'$ was \mbox{eventually} seen to \mbox{dominate} the precipitate structures. It is expected that this trend would \mbox{continue} until most Cu atoms have been drained from the solid solution. The results show that even a low Cu \mbox{content} ($0.01$ at.\%) can \mbox{significantly} affect the Al-Mg-Si \mbox{system} \mbox{precipitation}, \mbox{especially} during \mbox{overageing}.

The study demonstrates the potential of SPED when \mbox{combined} with other TEM techniques and data analysis \mbox{including} machine learning. The high information \mbox{content} in each SPED scan can \mbox{continually} be explored, and with the \mbox{advent} of powerful and dedicated machine learning \mbox{approaches} and improved detector \mbox{technologies}, this can potentially \mbox{uncover} further insights to the material microstructure.

\section*{Acknowledgements}
JKS, CDM and RH acknowledge support from the AMPERE project (247783), a Knowledge building Project for \mbox{Industry}, co-financed by The Research Council of Norway (RCN), and the industrial partners Norsk Hydro, Sapa, Gr\"anges, \mbox{Neuman} Aluminium Raufoss (Raufoss Technology) and \mbox{Nexans}. The (S)TEM work was carried out on the NORTEM \mbox{infrastructure} (NFR 197405) at the TEM Gemini Centre, Trondheim, \mbox{Norway}. 
All authors extend their gratitude to collaborating \mbox{researchers} Duncan N. Johnstone and prof. Paul A. Midgley from the \mbox{University} of Cambridge.

\section*{Data availability}
The TEM images presented in this work are given in raw/unprocessed form. The raw/processed scanning diffraction data required to reproduce these findings cannot be shared at this time as the data also forms part of an ongoing study.


\end{document}